%% file: ring-main.tex
 \documentclass[pre,showpacs,floats,twocolumn,floatfix,superscriptaddress]{revtex4-1}
\usepackage{lipsum}
\usepackage{mathrsfs}
\usepackage{bm,amsbsy,amssymb,amsmath}
\usepackage{caption}
\usepackage{subcaption}
\usepackage{graphics,graphicx,dcolumn,fleqn,epic,eepic,float,tabularx}
\usepackage{multirow,rotate,rotating,color}
\usepackage[utf8]{inputenc}
\newcommand{\figref}[1]{Fig.~\ref{fig:#1}}
\newcommand{\eqnref}[1]{Eq.~(\ref{eq:#1})} 
  \definecolor{tuered}{RGB}{214,0,74}
  \definecolor{tueblue}{RGB}{0,102,204}
  
  \newcommand{\revisedtext}[1]{\textcolor{black}{#1}}

\usepackage{tikz,pgfplots}
\usepackage{relsize}
\tikzset{fontscale/.style = {font=\relsize{#1}}}
\usetikzlibrary{calc}
\graphicspath{{img/}}
\begin{document}
\title{From dot to ring: the role of friction on the deposition pattern \\ of a drying colloidal suspension droplet}
 \author{Qingguang Xie}
  \email{q.xie1@tue.nl}
  \affiliation{Department of Applied Physics, Eindhoven University of Technology, P.O. Box 513, 5600MB Eindhoven, The Netherlands}
  \author{Jens Harting}
  \email{j.harting@fz-juelich.de}
  \affiliation{Helmholtz Institute Erlangen-N\"urnberg for Renewable Energy (IEK-11), Forschungszentrum J\"ulich, F\"urther Str. 248, 90429 N\"urnberg, Germany}
  \affiliation{Department of Applied Physics, Eindhoven University of Technology, P.O. Box 513, 5600MB Eindhoven, The Netherlands}
\date{\today}
\begin{abstract}
The deposition of particles on a substrate by drying a colloidal suspension
droplet is at the core of applications ranging from traditional printing on
paper to printable electronics or photovoltaic devices.  The self-pinning
induced by the accumulation of particles at the contact line plays an important
role in the formation of the deposition. In this paper, we investigate both
numerically and theoretically, the effect of friction between the particles and
the substrate on the deposition pattern.  Without friction, the contact line
shows a stick-slip behaviour and a dot-like deposit is left after the droplet
is evaporated. By increasing the friction force, we observe a transition from a dot-like
to a ring-like deposit. We propose a theoretical model to predict the effective
radius of the particle deposition as a function of the friction force. Our
theoretical model predicts a critical friction force when the self-pinning
happens and the effective radius of deposit increases with increasing friction
force, confirmed by our simulation results. Our results can find implications
for developing active control strategies for the deposition of drying droplets.
\end{abstract}

 \pacs{
  47.11.-j, 
  47.55.Kf, 
  77.84.Nh. 
  }
\maketitle
\input{introduction.tex}
\input{model.tex}
\input{results-puredrop.tex}
\input{results-mddrop.tex}
\input{results-discussion.tex}
 \input{conclusion.tex}
\begin{acknowledgments}
Financial support is acknowledged from the Netherlands Organization for
Scientific Research (NWO) through project 13291 and a NWO Industrial
Partnership Programme (IPP).  This research programme is co-financed by
Océ-Technologies B.V., University of Twente and Eindhoven University of
Technology. We thank the J\"ulich Supercomputing Centre and the High
Performance Computing Center Stuttgart for the technical support and allocated
CPU time.
 \end{acknowledgments}
\appendix 
\input{ring-main-supple.tex}

 \bibliographystyle{unsrt}
 \bibliography{biblio/thesis-ref.bib}

\end{document}

%% file: introduction.tex
\section{Introduction} 
 \label{sec:intro}
Drying a colloidal suspension droplet is common in many industrial
applications, such as inkjet printing~\cite{Singh2010} and coating
processes~\cite{Grainger1998} with applications ranging from the poduction of
newspapers and magazines to printed electronic or even photovoltaic devices.
During liquid evaporation, the particles accumulate and finally form a dry
solid deposit on the
substrate~\cite{Deegan1997,Han2012,Larson2014,Anyfantakis2015,Zhong2015}.  The
homogeneity and conformation of the deposition are crucial for the device
performance~\cite{Park2006,GIROTT2009}.  Therefore, a detailed microscopic
understanding and active control strategies for deposition formation process
are needed.

The particle deposition pattern is largely affected by the evaporation
behaviour of the droplet. \revisedtext{Generally, there are two basic
modes~\cite{Picknett1977,Cazabat2010, Detlef2015} of droplet
evaporation: constant contact radius (CR) mode, constant contact angle (CA)
mode.}  In the CR mode, the droplet evaporates with
constant radius while its contact angle decreases~\cite{Picknett1977}.  In the
CA mode, the droplet keeps a constant contact angle but the contact radius
decreases~\cite{Picknett1977}. \revisedtext{
In reality, the contact line
mostly shows combinations of the CA and CR modes due to irregular roughness of the substrate, which is called ``stick-slip mode''~\cite{Shanahan1995, Stauber2014}.} 
In the stick-slip mode, the droplet first evaporates in the CR mode until a certain contact angle is reached.  Then, the
evaporation mode switches to the CA mode~\cite{Cazabat2010}. The CA and CR
mode can be treated as extreme cases of the stick-slip mode.

Drying of a pure liquid drop on a solid substrate generally shows a stick-slip
contact line behaviour~\cite{Orejon2011,Pham2017}. The droplet starts to
evaporate with a pinned contact line, followed by a second phase where the
contact line depins, the contact radius shrinks and the contact angle remains
constant. Deegan~\cite{Deegan2000b} observed that the second phase can be
absent in drying colloidal drops. The contact line dynamics can be altered by
the accumulation of the solute at the contact line. A self-pinning mechanism is
proposed: the roughness or chemical heterogeneities of the substrate provide a
primary source of contact line pinning, then the accumulation of colloidal
particles at the contact line apparently strengthens the pinning, and
eventually becomes the dominant contribution to the contact line pinning. The
evaporation of a colloidal suspension droplet with pinned contact line
introduces an outward capillary flow inside the droplet, transporting the
colloidal particles to the edge~\cite{Deegan1997,Deegan2000}. The particles
accumulate at the contact line, strengthen the pinning of the contact line, and
finally form a ring-like deposit known as the
``coffee-ring''~\cite{Deegan1997}.

Recently, Sangani \textit{et al.}~\cite{Sangani2009} found that the capillary
force subjected on particles at the contact line can be sufficiently large to
overcome the self-pinning. The capillary force pushes the particles inward,
resulting in a dot-like deposit. Weon \textit{et al.}~\cite{Weon2013} further
studied the self-pinning effect of a spreading colloidal droplet. They found
that the capillary force is dominant for the self-pinning, while the friction
force between particle and substrate is negligible. However, very recently,
Wang \textit{et al.}~\cite{Wang2016} studied the effect of a single colloidal
particle on the contact line using the molecular dynamics method.  They found
that the contact line can experience pinning, depinning and the
pinning-depinning transition dependent on the friction force between particle
and substrate.  Meanwhile, Man and Doi~\cite{Doi2016} showed that the contact
line friction is a key parameter for a ring to mountain-like transition in the
deposition of drying droplets. 

In this paper, we numerically study the effect of friction between the
particles and the substrate on the deposition using the lattice Boltzmann
method. Firstly, we validate the applicability of our model by studying the CR
and stick-slip mode of an evaporating pure droplet.  We compare the time
dependent radial velocity of a drying pure droplet in CR mode with its
respective analytical prediction. Then, we investigate the particle deposition
obtained from a drying colloidal suspension droplet by varying the friction
force. Without friction, the contact line experiences a similar stick-slip mode
and particles form a dot-like deposit after the droplet is evaporated.  With
increasing friction force, the deposition shows a dot-like to ring-like
transition, indicating that the particles can introduce self-pinning for a
large friction force. 
\revisedtext{
In this work, we consider a perfectly flat substrate and 
demonstrate that the particles with large friction can act 
as topographical defects, which is consistent with the results in previous work~\cite{Pham2017}.
However, the transport of colloidal particles is usually modeled with a convection-diffusion equation~\cite{Espin2014,Pham2017}, 
and the interactions between particles and substrate are neglected.  
Here, we show that the interactions between particle and substrate can play an 
important role on the deposition pattern.}
We propose a theoretical model to predict the effective
radius of deposition as a function of the friction force, which is in  good
agreement with our simulation results. Moreover, our theoretical model reveals
that self-pinning is determined by the competition between capillary and
friction forces. 

%% file: model.tex
\section{Simulation Method}
\label{sec:method}
We use the lattice Boltzmann method (LBM) which can be treated as an
alternative numerical method to describe the dynamics of the fluid. In the
limit of small Knnudsen and Mach numbers, the Navier-Stokes equations are
recovered~\cite{Succi2001}. In the past two decades, the LBM has demonstrated
itself as a powerful tool for numerical simulations of fluid
flows~\cite{Succi2001,Raa04}, and has been extended to simulate
multiphase/multicomponent fluids~\cite{Shan1993,Liu2016} and suspensions of
particles of arbitrary shape and
wettability~\cite{ladd-verberg2001,Jansen2011,Gunther2013a,Xie2015}.  The LBM
is a local mesoscopic algorithm, allowing for efficient parallel
implementation. We review some relevant details in the following and refer the
reader to the relevant
literature~\cite{Jansen2011,Frijters2012,DennisXieJens2016,Liu2016} for a
detailed description of the method and our implementation.
 
We utilize the pseudopotential multicomponent LBM of Shan and Chen~\cite{Shan1993} 
with a D3Q19 lattice~\cite{Qian1992}.
Here, two fluid components are modelled 
by following the evolution of each distribution function 
discretized in space and time according to the lattice Boltzmann equation:
\begin{eqnarray}
  \label{eq:LBG}
  f_i^c(\mathbf{x} + \mathbf{e}_i \Delta t , t + \Delta t)&= &f_i^c(\mathbf{x},t) - \frac{\Delta t} {\tau^c} [  f_i^c(\mathbf{x},t) - \nonumber \\
  &&f_i^\mathrm{eq}(\rho^c(\mathbf{x},t),\mathbf{u}^c(\mathbf{x},t))]
  \mbox{,}
\end{eqnarray}
where $i=0,...,18$. $f_i^c(\mathbf{x},t)$ are the single-particle distribution
functions for fluid component $c=1$ or $2$, and $\mathbf{e}_i$ is the discrete
velocity in the $i$th direction.  $\tau^c$ is the relaxation time for component
$c$ and determines the viscosity.  The macroscopic densities and velocities for
each component are defined as  $ \rho^c(\mathbf{x},t) = \rho_0
\sum_if^c_i(\mathbf{x},t)$, where $\rho_0$ is a reference density, and
$\mathbf{u}^c(\mathbf{x},t) = \sum_i  f^c_i(\mathbf{x},t)
\mathbf{e}_i/\rho^c(\mathbf{x},t)$, respectively. 
Here, $f_i^\mathrm{eq}$ is the second-order equilibrium distribution function, defined as
\begin{eqnarray}
  \label{eq:eqdis}
  f_i^{\mathrm{eq}}(\rho^c,\mathbf{u}^c) &=& \omega_i \rho^c \bigg[ 1 + \frac{\mathbf{e}_i \cdot \mathbf{u}^c}{c_s^2} \nonumber \\
  && - \frac{ \left( \mathbf{u}^c \cdot \mathbf{u}^c \right) }{2 c_s^2} + 
  \frac{ \left( \mathbf{e}_i \cdot \mathbf{u}^c \right)^2}{2 c_s^4}  \bigg]
  \mbox{.}
\end{eqnarray}
where $\omega_i$ is a coefficient depending on the direction: $\omega_0=1/3$
for the zero velocity, $\omega_{1,\dots,6}=1/18$ for the six nearest neighbors
and $\omega_{7,\dots,18}=1/36$ for the nearest neighbors in diagonal direction.
$c_s = \frac{1}{\sqrt{3}} \frac{\Delta x}{\Delta t}$ is the speed of sound.

For convenience we choose the lattice constant $\Delta x$, the timestep $
\Delta t$, the unit mass $\rho_0 $ and the relaxation time $\tau^c$ to be
unity, which leads to a kinematic viscosity $\nu^c$ $=$ $\frac{1}{6}$ in
lattice units.

The pseudopotential multicomponent model introduces a mean-field interaction force 
\begin{equation}
  \label{eq:sc}
\!\!\!\!\!  \mathbf{F}^c(\mathbf{x},t) = -\Psi^c(\mathbf{x},t) \sum_{\bar{c}} \sum_{i} \omega_i g_{c\bar{c}} \Psi^{\bar{c}}(\mathbf{x}+\mathbf{e}_i,t) \mathbf{e}_i
\end{equation}
between fluid components $c$ and $\bar{c}$~\cite{Shan1993}, 
in which $g_{c\bar{c}}$ is a coupling constant, eventually leading to a demixing of the fluids.
We note $\gamma_{12}$ as the surface tension of the interface.
$\Psi^c(\mathbf{x},t)$ is an ``effective mass'', chosen as the functional form
\begin{equation}
  \label{eq:psifunc}
  \Psi^c(\mathbf{x},t) \equiv \Psi(\rho^c(\mathbf{x},t) ) = 1 - e^{-\rho^c(\mathbf{x},t)}
   \mbox{.}
\end{equation}

This force $\mathbf{F}^c(\mathbf{x},t)$ is then applied to the component $c$ by adding a 
shift $\Delta \mathbf{u}^c(\mathbf{x},t) =\frac{\tau^c \mathbf{F}^c(\mathbf{x},t)}{\rho^c(\mathbf{x},t)}$ 
to the velocity $\mathbf{u}^c(\mathbf{x},t)$ in the equilibrium distribution.

We introduce an interaction force between the fluid and wall, inspired by the work of Huang et al.~\cite{Huang2007}
\begin{equation}
\mathbf{F}^c(\mathbf{x}) =  - g^{wc} \Psi^c (\mathbf{x}) \sum_{i} \omega_{i} s(\mathbf{x} + \mathbf{e}_{i}) \mathbf{e}_{i}
\mbox{,}
\label{eq:rhowall2}
\end{equation}
where $g^{wc}$ is a constant.  Here, $s(\mathbf{x} + \mathbf{e}_{i})= 1$ if
$\mathbf{x} + \mathbf{e}_{i}$ is a solid lattice site, and $s(\mathbf{x} +
\mathbf{e}_{i})=0$ otherwise.

When the interaction parameter $g_{c\bar{c}}$ in~\eqnref{sc} is
properly chosen~\cite{Martys1996a}, a separation of components takes place.
Each component separates into a denser majority phase of density
$\rho_{ma}$ and a lighter minority phase of density $\rho_{mi}$,
respectively. \revisedtext{The interface is diffusive, which avoids 
the stress singularity at the moving contact line usually occurring in sharp-interface models.}

In order to trigger evaporation, we impose the density of component $c$
at the boundary sites $\mathbf{x}_H$ to be of constant value 
$\rho^c(\mathbf{x}_H,t)= \rho^c_H$ by setting the distribution function of
component $c$ to~\cite{DennisXieJens2016} 
\begin{equation}
    f_i^c(\mathbf{x}_H,t) = f_i^\mathrm{eq}\left(\rho_H^c,\mathbf{u}^c_H(\mathbf{x}_H,t)\right),
\end{equation}
in which $\mathbf{u}^c_H(\mathbf{x}_H,t)=0$. In the case where the set density
$\rho_H^c$ is lower than the equilibrium minority density $\rho_{mi}^c$, a
density gradient in the vapor phase of component $c$ is formed. This gradient
causes component $c$ to diffuse towards the evaporation boundary.
\revisedtext{
We note that our evaporation model is diffusion dominated, 
which is validated in our previous work~\cite{DennisXieJens2016}.
For further discussions on different approaches (such as transfer-rate dominated) to model evaporation, 
we refer the reader to the
relevant literature~\cite{Murisic2011, Pham2017}.}

The colloidal particles are discretized on the fluid lattice and coupled to the fluid species by means of 
a modified bounce-back boundary condition as pioneered by Ladd and Aidun~\cite{ladd-verberg2001, AIDUN1998}.
The particles follow classical equations of motion 
\begin{eqnarray}
 \mathbf{F}_{\mathrm{p}}=m\cdot \frac{d\mathbf{u}_{\mathrm{p}}}{dt} \mbox{,} 
\label{eq: newton}
\end{eqnarray}
in which $\mathbf{F}_{\mathrm{p}}$ is the total force imposed on a particle
with mass $m$ and $\mathbf{u}_{\mathrm{p}}$ is the velocity the particle. The
trajectory of a colloidal particle is updated using a leap-frog integrator.

We fill the outer shell of the particle with a ``virtual'' fluid by an amount
$\Delta \rho$~\cite{Jansen2011, Frijters2012},
\begin{eqnarray}
 \rho_{\mathrm{virt}}^1(\mathbf{x},t)  &=& \overline{\rho}^1(\mathbf{x},t) + |\Delta\rho_p| \mbox{, } \\
 \rho_{\mathrm{virt}}^2(\mathbf{x},t) & =& \overline{\rho}^2(\mathbf{x},t) - |\Delta\rho_p|  \mbox{, }
 \label{eq:md_colour}
\end{eqnarray}
where $\overline{\rho}^1(\mathbf{x},t)$ and $\overline{\rho}^2(\mathbf{x},t)$
are the averages of the density of neighbouring fluid nodes for component $1$
and $2$, respectively.  The parameter $\Delta\rho_p$ is called the ``particle
colour'' and dictates the contact angle of the particle.  A particle colour
$\Delta\rho_p = 0$ corresponds to a contact angle of $\theta = 90^{\circ}$,
i.e. a neutrally wetting particle. 

\revisedtext{The momentum exchange between particles and fluid recovers the hydrodynamic forces, such as drag and lift forces. 
Moreover, our model recovers the lubrication interactions correctly when the distance 
between two particles is at least one lattice site.
If there is less than one lattice site between the particles, a lubrication correction is
introduced~\cite{ladd-verberg2001,JHT16,KHV10},
\begin{equation}
 \mathbf{F}_{ij} = -\frac{3\pi \mu a^2}{2} \hat{\mathbf{r}}_{ij} \hat{\mathbf{r}}_{ij} \cdot (\mathbf{u}_i-\mathbf{u}_j) \left( \frac{1}{r_{ij}-2a}-\frac{1}{\Delta_c} \right)
 \end{equation}
where $a$ is the radius of the particle, $\mathbf{\hat{r}}_{ij}=\frac{\mathbf{r}_{i}-\mathbf{r}_{j}}{|\mathbf{r}_{i}-\mathbf{r}_{j}|}$ 
is a unit vector pointing from one particle center to the other one and $r_{ij}$ is the distance between particle $i$ and $j$.
$\mathbf{u}_i$ and $\mathbf{u}_j$ are particle velocities. $\Delta_c$ is a constant and is chosen to be $\Delta_c=2/3$. 
}

To avoid overlapping of particles, we add a Hertz potential between the particles 
with the following form for two spheres with identical radii $a$~\cite{hertz1881}:
\begin{equation}
  \label{eq:hertz-potential}
  \phi_H = \left\{\begin{matrix}
K_H(2a-r_{ij})^{\frac{5}{2}}\quad\mbox{for}\quad r_{ij}\le 2a
\\ 
0, \quad\mbox{otherwise} 
\end{matrix}\right.
\end{equation}
Here, $K_H$ is the force constant and is chosen to be $K_H = 100$.

\revisedtext{For the interactions between particles and a substrate, 
the lubrication forces between particles and walls are modeled in a similar way as that between particles.
Additionaly,} we implement the
Lennard-Jones (LJ) potential between particles and a substrate as
\begin{equation}
\!\!\!\!\!\! \phi_{LJ} = 4\epsilon \left( (\frac{\sigma}{r_{iw}})^{12} - (\frac{\sigma}{r_{iw}})^6 \right) \quad\mbox{for}\quad r_{iw}\le 2.5a
\end{equation}
where $\epsilon$ is the depth of the potential well, $\sigma$ is the finite
distance at which the inter-particle potential is zero, and $r_{iw}$ is the
distance between a particle center and the substrate surface.  We set $\sigma$
equal to the particle radius $a$ in all simulations.

To model \revisedtext{the} tangential \revisedtext{friction} between particles and a substrate,
we apply a constant friction force on the particles as follows:
\begin{eqnarray}
\!\!\!\!\!\!\!\!\!\mathbf{F}_{f} = \frac{-F_f\mathbf{u}_{\parallel p}}{|\mathbf{u}_{\parallel p}|} \quad\mbox{for}\quad r_{iw}\le a+1 \quad \& \quad |\mathbf{F}_{\parallel p}| > F_f \mbox{,} \nonumber 
\\
\label{eq:friction}
\end{eqnarray}
where $\mathbf{u}_{\parallel p}$ is the component of velocity of the particle
parallel to the substrate.  Here, $\mathbf{F}_{\parallel p}$ is the force acting on the
particle due to the particle-particle interaction and particle-fluid
interactions in the direction parallel to the substrate. The constant $F_{f}$
represents the maximum friction.  When $|\mathbf{F}_{\parallel p}| \leq F_f$ and
$r_{iw}\le a+1$, the particles get stuck on the substrate. For simplicity, we assume that the static and kinetic friction are equal and the friction force is
independent on the normal force exerted on the particle. 
\revisedtext{
We note that~\eqnref{friction} satisfies the classical Coulomb law of friction, i.e.
the direction of the friction force is opposite to the sliding
velocity of the particles and the kinetic friction is independent on the sliding velocity.
Similar friction models have been successfully applied to investigate the structure formation of
a drying colloidal suspension film~\cite{FUJITA2015}
and granular dynamics~\cite{Schwager2005}.
Recently, Guo et al. experimentally measured the
friction between a nanoparticle and a substrate and demonstrated that the sliding kinetic friction is constant~\cite{Guo2013}.
}

\input{draw-drop.tex}

%% file: draw-drop.tex
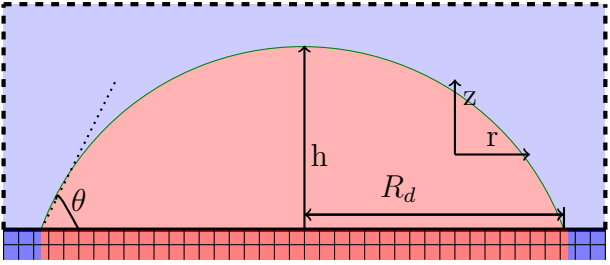
\begin{figure}[h!]
\centering
 \begin{tikzpicture}
	\fill[blue,opacity=0.2] (-4,0) rectangle (4.0,3.0);
	
	 \filldraw[fill=red!30!white, draw=green!50!black]
      (0.0,0) -- (3.45,0.0) arc (20:160:3.7) -- (-0.0,0);
	
	\draw[ultra thick,black] (-4.0,0) to (4.0,0);
	\draw[ultra thick,black, dashed] (-4.0,0) to (-4.0,3);
	\draw[ultra thick,black, dashed] (4.0,0) to (4.0,3);
	\draw[ultra thick,black, dashed] (-4.0,3) to (4.0,3);


 \draw[thick, black,arrows=->] (2,1) to (2,2);
 \node at (2.2,1.75) [fontscale=2] {z};
 \draw[thick, black,arrows=->] (2,1) to (3,1);
 \node at (2.5,1.2) [fontscale=2] {r};
  \draw[thick, black,arrows=->] (0,0) to (0,2.45);
 \node at (0.2,1.0) [fontscale=2] {h};
  \draw[thick, black,arrows=->] (0,0.2) to (3.45,0.2);
   \draw[thick, black,arrows=->] (3.45,0.2) to (0,0.2);
   \draw[thick, black] (3.45,0.0) to (3.45,0.3);
 \node at (1.25,0.55) [fontscale=2] {$R_d$};
  \draw[thick, black, dotted] (-3.5,0) to (-2.5,2);
  \draw[thick,black] (-3.0,0.0) to [out=120,in=90] (-3.3,0.4);
 \node at (-3.0,0.4) [fontscale=2] {$\theta$};
 
  \fill[red,opacity=0.5] (-3.5,-0.4) rectangle (3.5,0.0);
  \fill[blue,opacity=0.5] (-4.0,-0.4) rectangle (-3.5,0.0);
  \fill[blue,opacity=0.5] (3.5,-0.4) rectangle (4.0,0.0);
  \draw[step=2mm, black] (-4.0,-0.4) grid (4.0,0.0);
\end{tikzpicture}

\caption{Sketch of a droplet sitting on a substrate and covered by another fluid.
The droplet height at $r=0$ is $h$, the contact radius is $R_d$ and the contact angle is $\theta$.
We place a chemically patterned substrate with variable wettability at the bottom, while the boundaries normal to the substrate are periodic. 
After equilibration, we apply evaporation boundary conditions at the
sides and the top of the system (denoted by the dashed lines).
}
\label{fig:drop-geo-fluid}
\end{figure}

%% file: results-puredrop.tex
\section{Results and discussion}
\subsection{Pure droplet}
We start out to simulate the evaporation of a pure droplet sitting on a solid
substrate, as illustrated in~\figref{drop-geo-fluid}. The simulations utilize a
system size of $256\times 256 \times 144$ and the substrate is chemically
patterned with a variable wettability: a superhydrophilic circle ($\theta
\approx 0^{\circ}$) of radius $R_s=115$ is located at the center surrounded by a
superhydrophobic area ($\theta \approx 180^{\circ}$). We initialize the droplet
with a contact radius $R_0=115$ of initial maximal height $h_0=100$, and
densities $\rho^c_{ma} = \rho^{\bar{c}}_{ma}=0.70$, $\rho^c_{mi} =
\rho^{\bar{c}}_{mi}=0.04$. The initial contact angle of the droplet is
$\theta_0 \approx 82^{\circ}$.  After equilibration, we apply evaporation
boundary conditions with $\rho_{H}^{c}=0.01$ at the sides and the top of the
system, as shown by the dashed lines in~\figref{drop-geo-fluid}.
\begin{figure}[t!]
\centering
 \includegraphics[width=0.4\textwidth]{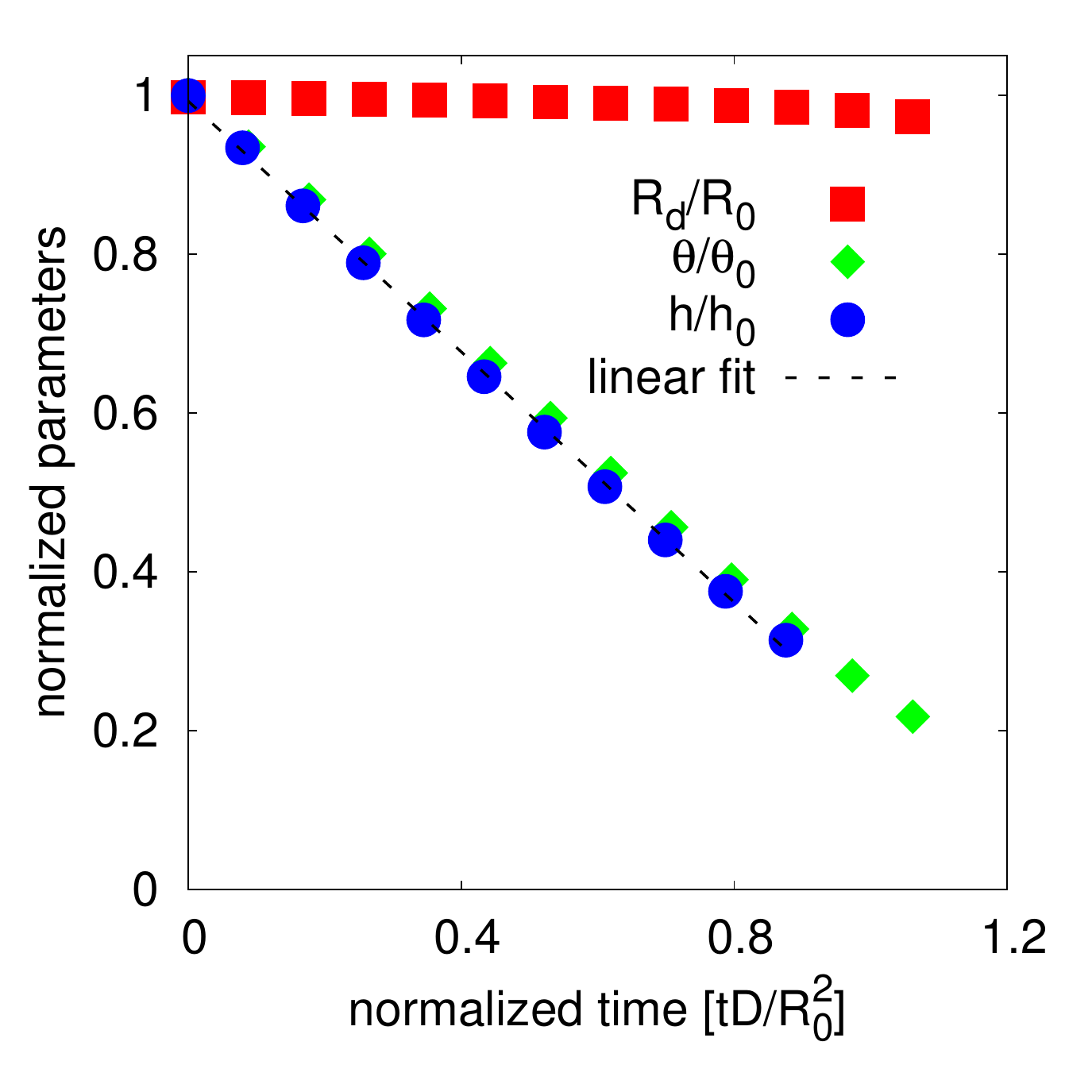}
 \caption{The normalized contact radius $R_d/R_0$, droplet height $h/h_0$ and
contact angle $\theta/\theta_0$ of the droplet as a function of time for
$\rho_{H}^{c} = 0.01$.  The symbols are simulation data and the dashed line is
a fitted linear function.  The droplet height decreases linearly, which agrees
qualitatively with the theoretical prediction~\eqnref{htime}. 
}
 \label{fig:basetheta-t0}
\end{figure}

\figref{basetheta-t0} shows the evolution of the normalized contact radius
$R_d/R_0$, the normalized droplet height $h/h_0$, and the normalized contact
angle $\theta/\theta_0$ versus time. The time is normalized with the
diffusivity of the fluid, $D\approx 0.117$~\cite{DennisXieJens2016}.  During
the evaporation, the contact angle of the droplet decreases while the contact
radius keeps constant. This is the so-called CA mode~\cite{Picknett1977},
resulting from the contact line pinning caused by contact angle heterogeneities
of the substrate. The droplet height and contact angle decrease linearly.

\begin{figure}[t!]
\centering
 \includegraphics[width=0.4\textwidth]{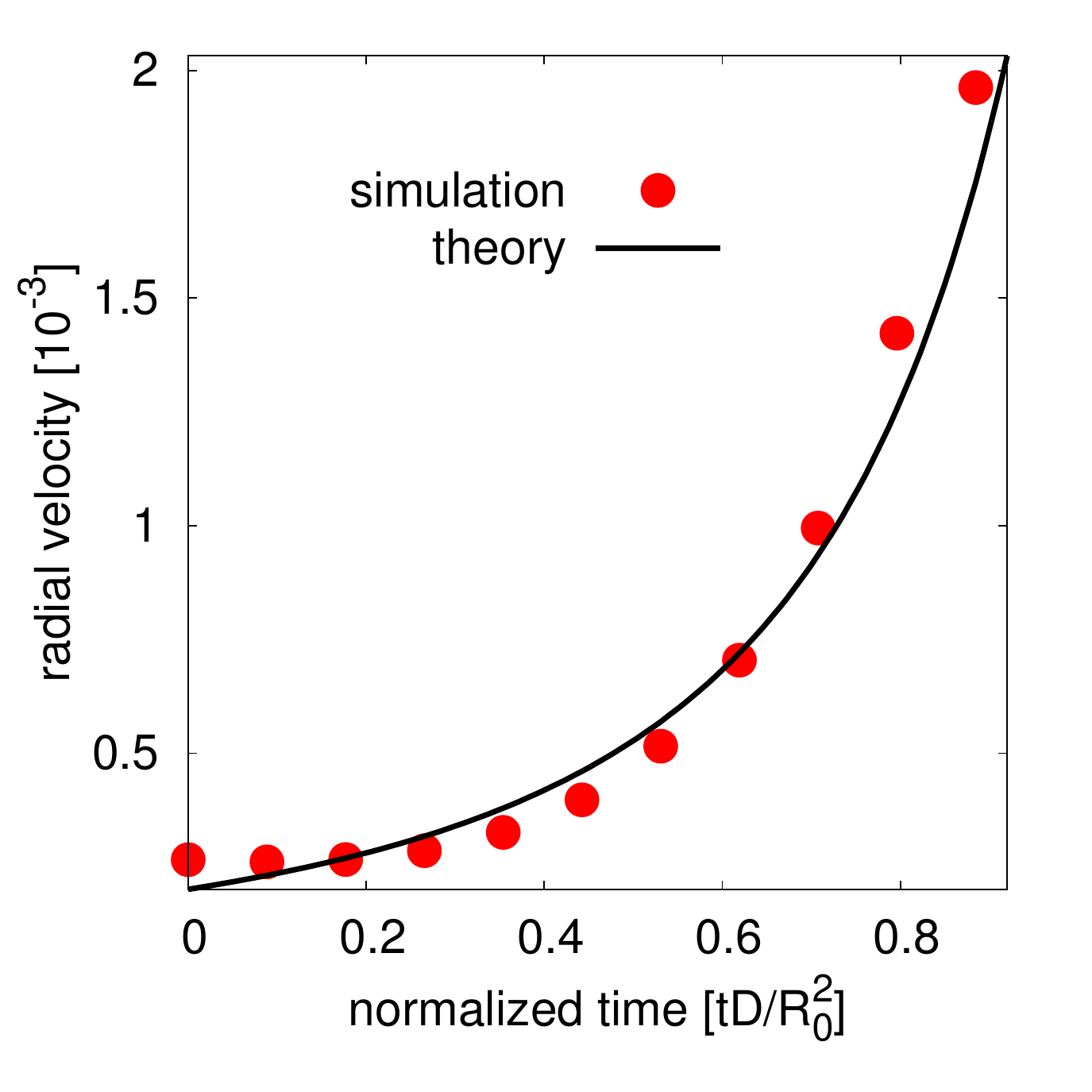}
 \caption{The time evolution of the radial velocity at position $r=100$, $z=6$ 
 for $\rho_{H}^{c}=0.01$. 
 The simulation data (symbols) is in a good agreement 
 with the theoretical prediction~\eqnref{radiav} (solid line).}
 \label{fig:v-time}
\end{figure}
The evaporation of a droplet with pinned contact line introduces
an outward capillary flow inside the droplet~\cite{Deegan1997,Popov2005}. 
Following the principle of mass conservation~\cite{Marin2011,Deegan1997,Popov2005} and assuming an infinite size of the system, 
the evolution of the droplet height in the limit of small contact angles is obtained as
\begin{equation}
 h(0,t) =  \frac{8D \Delta \rho}{\pi \rho R_d} (t_e-t)
 \mbox{,}
 \label{eq:htime}
\end{equation}
with $t_e$ denoting the total life time of the droplet. We note that~\eqnref{htime} predicts 
that the droplet height decreases linearly with time, 
which is consistent with our simulation results shown in~\figref{basetheta-t0}.
Using the lubrication approximation~\cite{Marin2011}, 
one can estimate the radial velocity near the substrate as 
\begin{equation}
 u(r,z,t) = \frac{3}{h^2(r,t)}  \bar{u} \left(h(r,t)z-\frac{1}{2}z^2 \right)
 \mbox{,}
 \label{eq:radiav}
\end{equation}
where $\bar{u}$ is the average radial velocity and is given as
\begin{equation}
 \bar{u} = \frac{R_{d}^{3}}{4r} \left[ \frac{1}{\sqrt{R_{d}^{2}-r^2}}  - \frac{1}{R_{d}^{3}}(R_{d}^{2}-r^2) \right] \frac{1}{t_e-t}
 \label{eq:velave}
 \mbox{.}
\end{equation}
The term $\frac{1}{t_e-t}$ in~\eqnref{velave} predicts that the radial velocity
diverges near the end of the lifetime of the droplet, which is confirmed by
experiments~\cite{Marin2011}.  For a detailed derivation, we refer the reader
to the Appendix A. 

In the simulations, we measure the radial velocity at position $r=100,z=6$ and
compare the simulation data with the theoretical analysis~\eqnref{radiav} in
\figref{v-time}.  The total lifetime of the droplet $t_e$ is determined from a
linear fit of the time evolution of the droplet height (dashed line shown
in~\figref{basetheta-t0}).  The radial velocity increases with time and shows a
rapid increase at later times, which agrees well with the theoretical
prediction~\eqnref{radiav}.
 
\begin{figure}[h!]
\centering
 \includegraphics[width=0.4\textwidth]{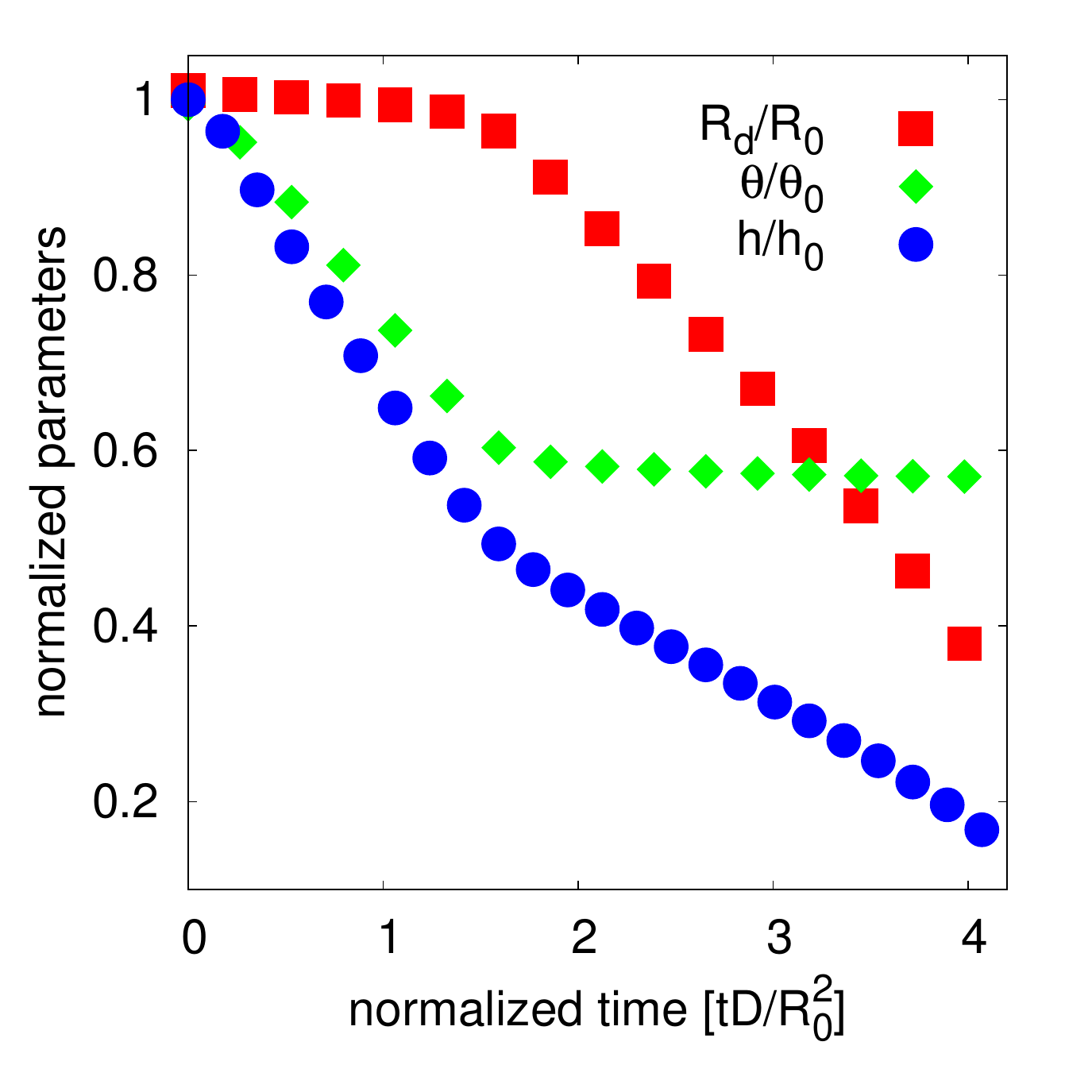}
 \caption{The evolution of the contact radius $R_d/R_0$, the droplet height $h/h_0$, and 
 the contact angle $\theta/\theta_0$ of a pure droplet for $\rho_{H}^{c} = 0.01$. 
 The contact line shows a stick-slip behavior: firstly, the contact angle decreases while 
 the contact radius keeps constant, followed by a phase where 
 the contact angle keeps constant while the contact radius decreases.
 } 
 \label{fig:basetheta-t1}
\end{figure}
Next, we investigate the stick-slip mode of a drying droplet by creating a
hydrophilic circular area with contact angle $\approx 56^{\circ}$ located at
the center of the substrate and the remaining area is hydrophobic of contact
angle $\approx 124^{\circ}$.  We place a droplet with an initial radius
$R_0=115$ and an initial height $h_0=115$, therefore, the initial contact angle
is $\theta_0=90^{\circ}$.

\figref{basetheta-t1} shows the time evolution of the contact radius $R_d/R_0$,
the droplet height $h/h_0$, and the contact angle $\theta/\theta_0$ of the
evaporating droplet.  In the beginning, the contact angle decreases while the
contact radius keeps constant. This indicates that the contact line is pinned
at the border between the hydrophilic and the hydrophobic area due to the
wettability heterogeneities of the substrate.  When the contact angles reaches
$\approx 56^{\circ}$, it stays constant, while the contact radius begins to
decrease, indicating that a depinning of the contact line occurs.  The droplet
height decreases throughout the drying time of the droplet, but with a faster
decreasing rate in the pinned phase than in the unpinned phase, which is
consistent with experimental results~\cite{Detlef2015}.  Therefore, we
demonstrate that our system is able to reproduce the stick-slip mode accurately
by utilizing the patterned wettability of the substrate. \revisedtext{
We note that the stick-slip motion can also be triggered 
by the topological features on the substrate~\cite{Pham2017}. 
For simplicity, we initialize the substrate 
with patterned wettability to introduce the stick-slip mode.}

%% file: results-mddrop.tex
\subsection{Colloidal suspension droplet}
In the following we investigate the evaporation of a colloidal suspension
droplet.  We initialize the droplet with a particle volume concentration
$\Phi\approx 0.89\% $ which corresponds to $500$ particles of radius $a=3$.
The particles are slightly hydrophilic with a contact angle $\theta_p \approx
75^{\circ}$ and the substrate has a hydrophilic circular area with a contact
angle $\approx 56^{\circ}$ located at the center surrounded by a hydrophobic
area of contact angle $\approx 124^{\circ}$. We note that the pure droplet
shows a stick-slip drying mode on this specific substrate.  We apply a constant
friction force with magnitude $F_f$ following \eqnref{friction} between
particles and substrate.  Then, we let the system equilibrate and apply
evaporation boundary conditions on the sides and the top of the system.

\begin{figure}[h!]
\centering
 \includegraphics[width=0.4\textwidth]{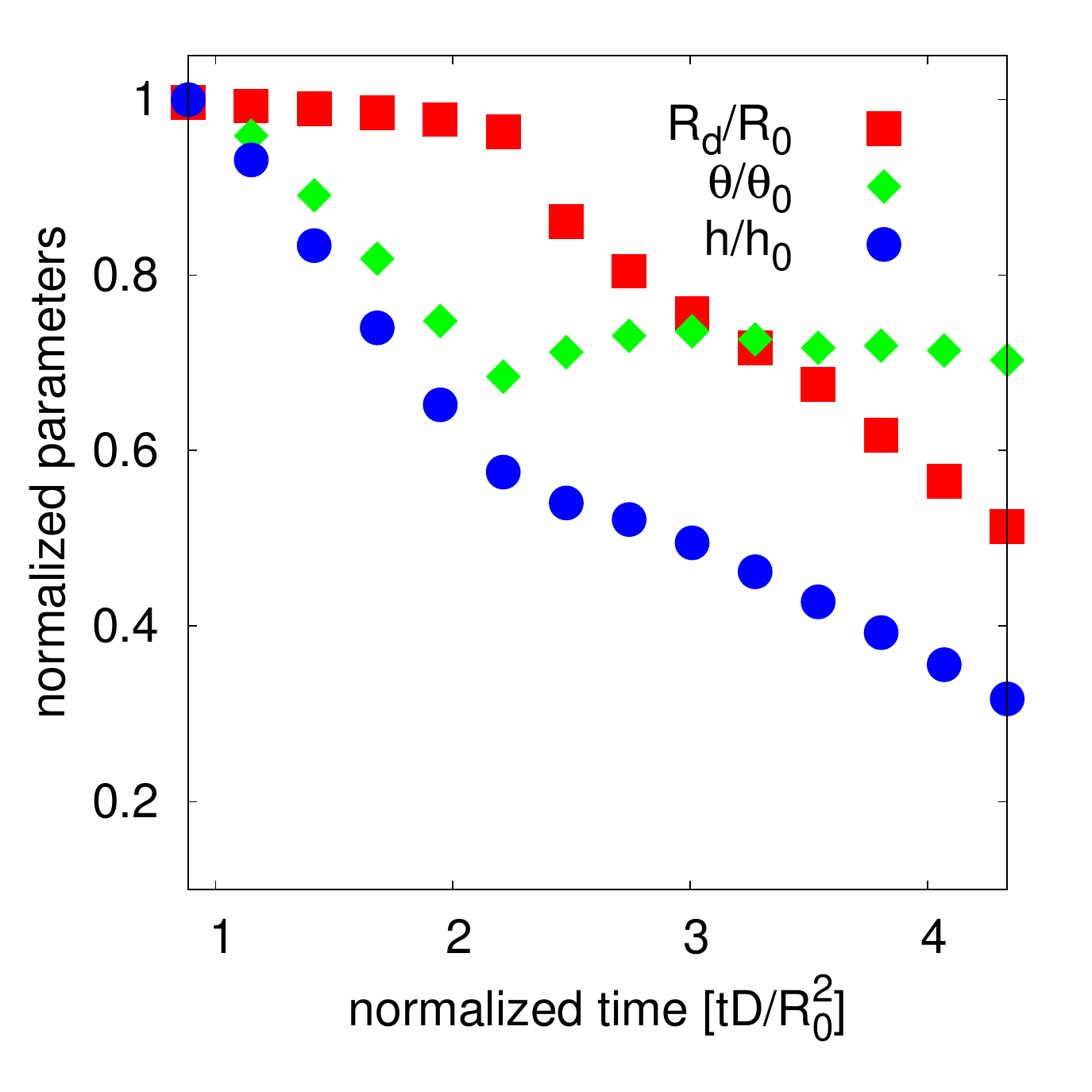}
 \caption{The evolution of the contact radius $R_d/R_0$, the droplet height $h/h_0$ and 
 the contact angle $\theta/\theta_0$ of a colloidal suspension droplet without friction.}
 \label{fig:bch-smallfri}
\end{figure}
Firstly, we set the friction force to zero and investigate the contact line dynamics. 
\figref{bch-smallfri} shows that the contact line of the colloidal suspension droplet 
initially follows the constant radius mode: the contact radius keeps constant and 
the contact angle decreases.
\begin{figure}[h!]
\begin{subfigure}{.23\textwidth}
\includegraphics[width= 0.95\textwidth]{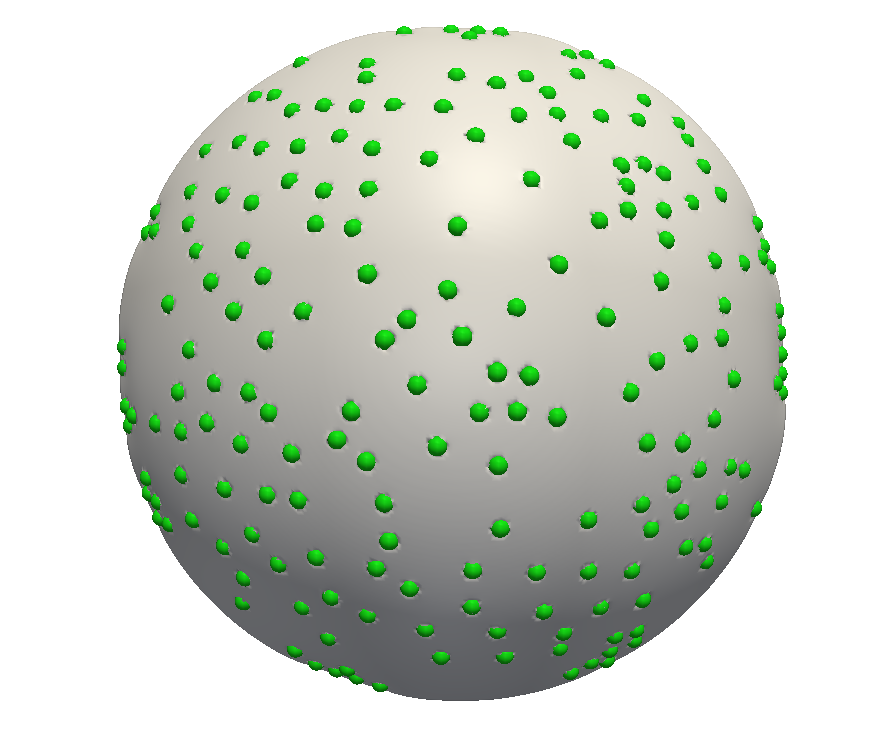}
 \subcaption{$tD/R_{0}^{2}=1.2$}
 \label{fig:frs-1}
 \end{subfigure}
 \begin{subfigure}{.23\textwidth}
\includegraphics[width= 0.95\textwidth]{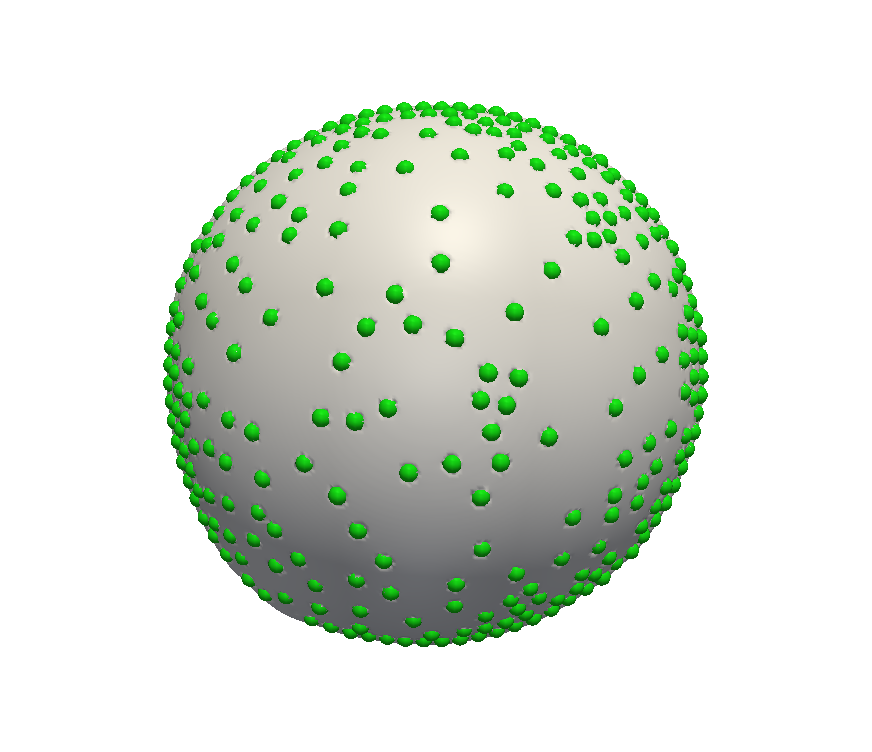}
 \subcaption{$tD/R_{0}^{2}=2.1$}
 \label{fig:frs-2}
 \end{subfigure}
 \\
 \begin{subfigure}{.23\textwidth}
\includegraphics[width= 0.95\textwidth]{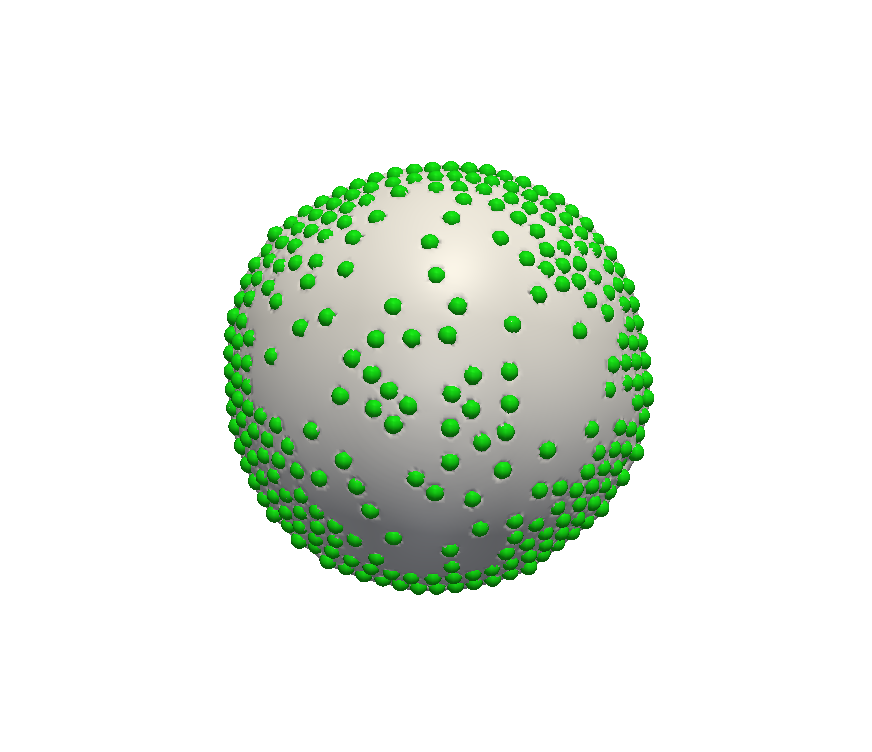}
 \subcaption{$tD/R_{0}^{2}=3.0$}
 \label{fig:frs-3}
 \end{subfigure}
  \begin{subfigure}{.23\textwidth}
\includegraphics[width= 0.95\textwidth]{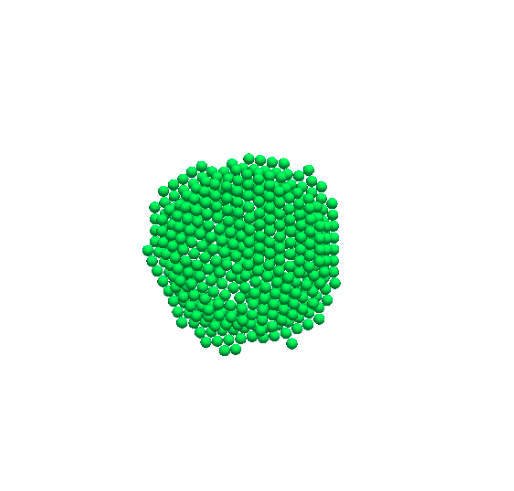}
 \subcaption{$tD/R_{0}^{2}=4.1$}
 \label{fig:frs-4}
 \end{subfigure}
 \caption{Snapshots of a drying colloidal suspension droplet without friction.
The particles are transported to the contact angle, and are pushed inwardly by
the capillary force. Finally, the particles form a dot-like pattern.}
\label{fig:frs}
 \end{figure}
When the contact angle reaches $\theta \approx 56^{\circ}$, it slightly
increases and then stays constant.  Meanwhile, the contact radius begins to
decrease, indicating that the contact line depins.  The droplet height keeps
decreasing in the whole process of drying, which is also observed in the
stick-slip mode of a drying pure droplet.  We note that the contact angle at
the slip stage ($\theta \approx 75^{\circ}$) is larger than the one observed
for a pure droplet ($\theta \approx 56^{\circ}$) (\figref{basetheta-t1}).  The
reason is that the particles accumulate at the contact line and serve as a new
substrate.  The contact line is pinned at the particle surface. Thus, the
contact angle of the droplet is determined by the particle wettablity, i.e.
$\theta_p \approx 75^{\circ}$.

We show the corresponding drying process of the colloidal suspension droplet
without friction in~\figref{frs}. In the initial CR mode of evaporation, the
particles are transported to the contact line and intend to pin the contact
line (\figref{frs-1}).  When the contact angle reaches a critical value, the
contact line depins, recedes and pulls the particles inwards by the capillary
force (\figref{frs-2}).  Then, the contact line transits to the constant angle
mode and the particles keep accumulating at the contact line (\figref{frs-3}).
The capillary force keeps pulling particles inward, and finally they form a
dot-like pattern (\figref{frs-4}).

\begin{figure}[h!]
\centering
 \includegraphics[width=0.4\textwidth]{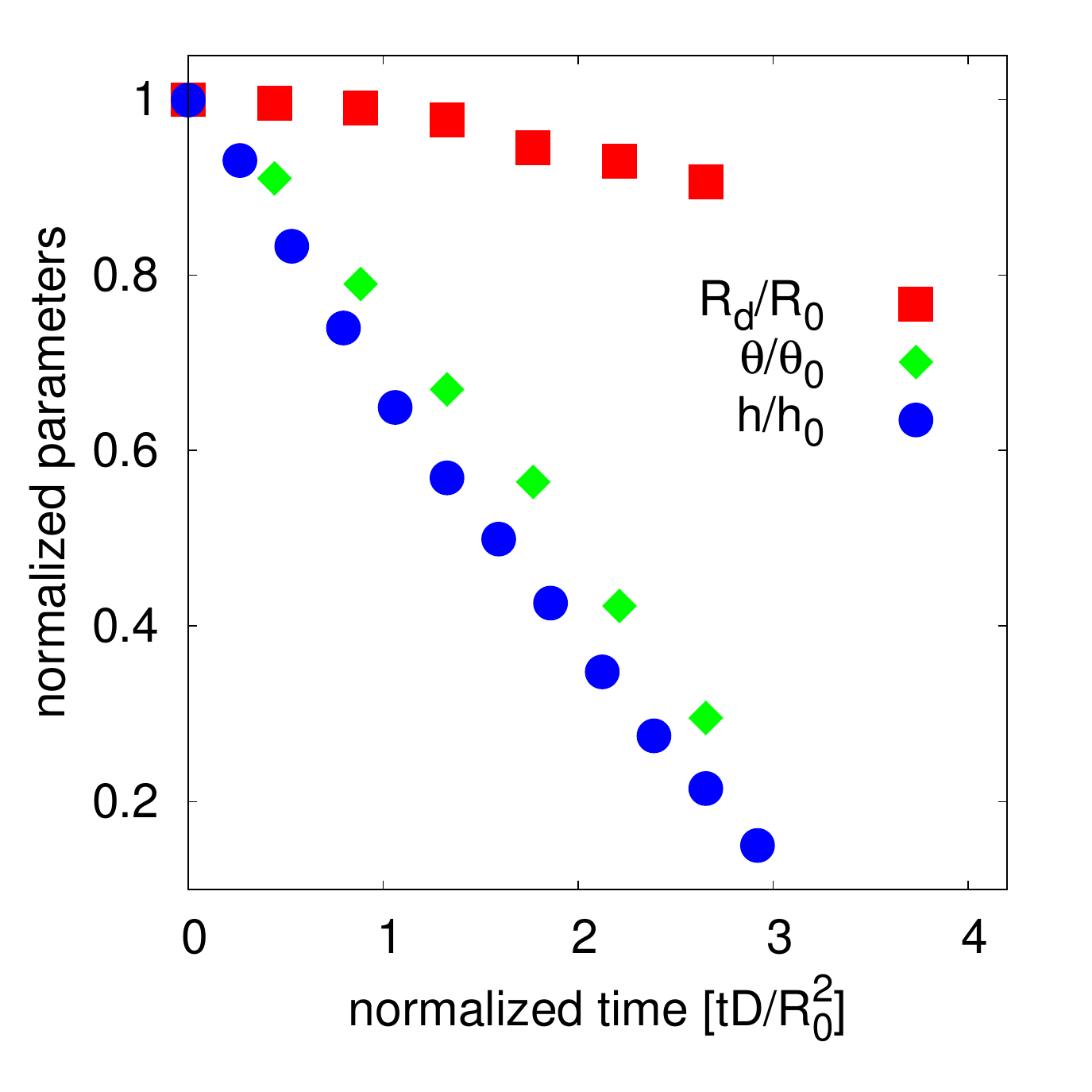}
 \caption{The evolution of the contact radius $R_d/R_0$, the droplet height
$h/h_0$ and the contact angle $\theta/\theta_0$ of an evaporating colloidal
suspension droplet for a large friction force $F_f/a \gamma_{12} = 0.07$.}
 \label{fig:bch-bigfri}
\end{figure}

In order to investigate the influence of friction between particles and
substrate, we set $F_f/(a\gamma_{12})=0.07$. 
  \figref{bch-bigfri} shows the
time evolution of the contact radius $R_d/R_0$, the droplet height $h/h_0$ and
the contact angle $\theta/\theta_0$ of the evaporating colloidal suspension
droplet.  At first, the contact radius keeps constant and the contact angle
decreases, which is similar to the behavior of a drying colloidal suspension
droplet without friction, as shown in~\figref{bch-smallfri}.  Interestingly,
when the contact angle decreases to $\approx 56^{\circ}$, there is only a
slight decrease of the contact radius, in contrast to the continuous decrease
of the contact radius observed in the drying of a pure droplet and the drying
of a colloidal suspension droplet without friction.  Moreover, the contact
angle keeps decreasing, which is similar to the constant radius mode observed
for a drying pure droplet (\figref{basetheta-t0}).

\begin{figure}[h!]
\begin{subfigure}{.23\textwidth}
\includegraphics[width= 0.90\textwidth]{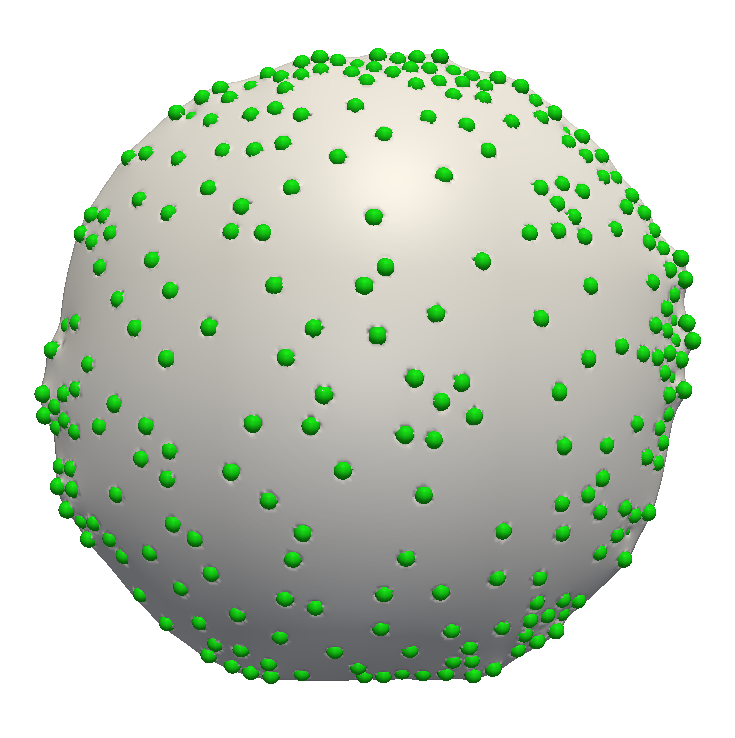}
 \subcaption{$tD/R_{0}^{2}=1.7$}
 \label{fig:frb-1}
 \end{subfigure}
 \hspace{0mm}
 \begin{subfigure}{.23\textwidth}
\includegraphics[width= 0.90\textwidth]{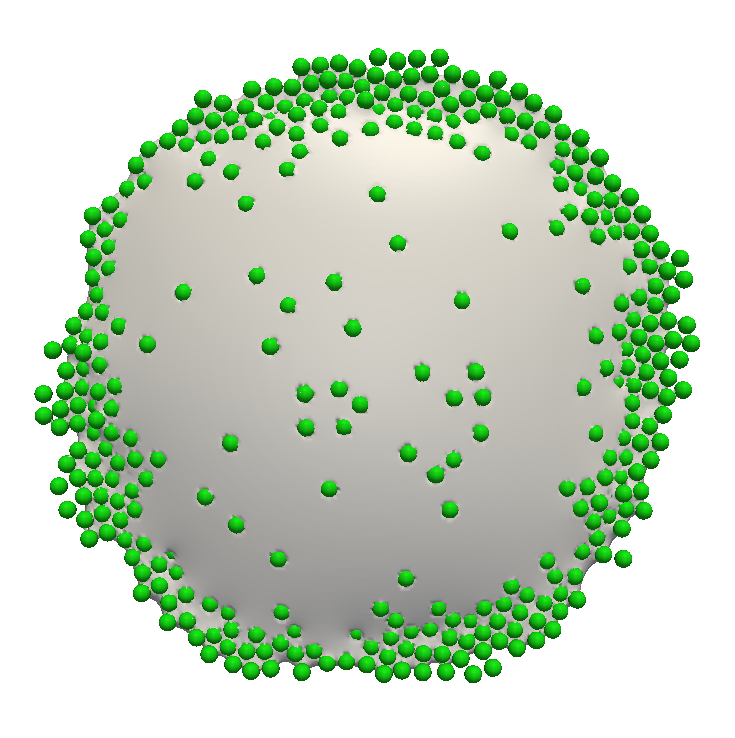}
 \subcaption{$tD/R_{0}^{2}=2.7$}
 \label{fig:frb-2}
 \end{subfigure}
 \\
 \begin{subfigure}{.23\textwidth}
\includegraphics[width= 0.90\textwidth]{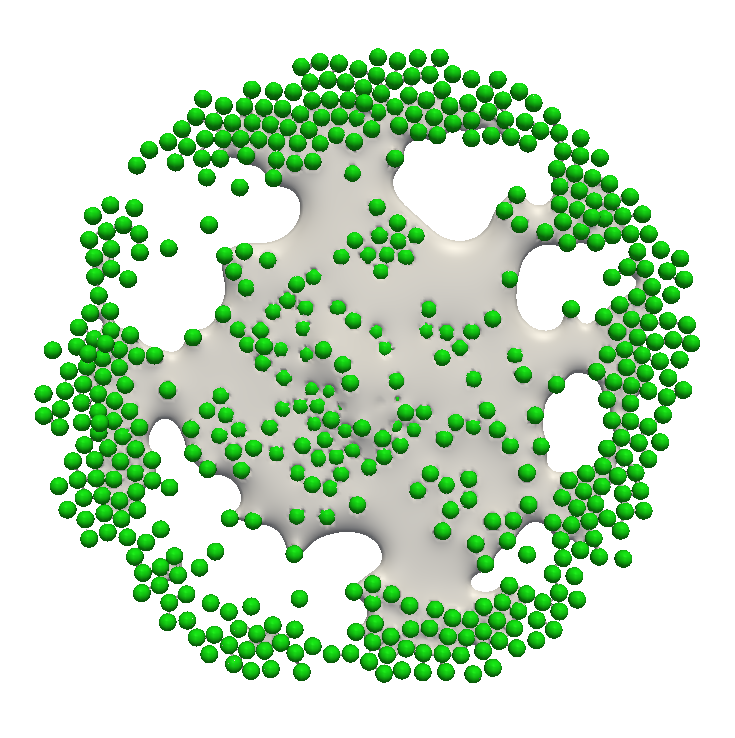}
 \subcaption{$tD/R_{0}^{2}=3.4$}
 \label{fig:frb-3}
 \end{subfigure}
  \hspace{0mm}
  \begin{subfigure}{.23\textwidth}
\includegraphics[width= 0.90\textwidth]{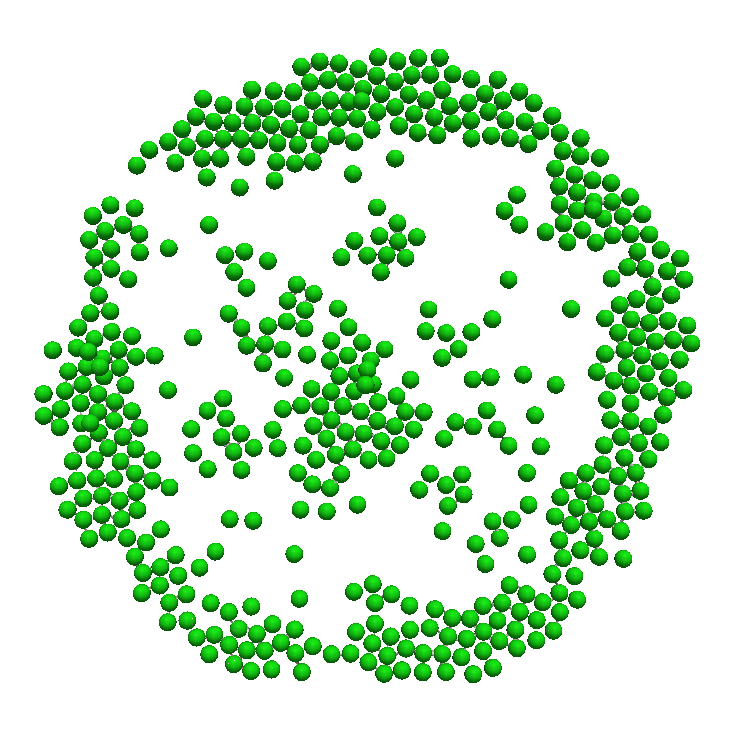}
 \subcaption{$tD/R_{0}^{2}=4.0$}
 \label{fig:frb-4}
 \end{subfigure}
 \caption{Snapshots of the drying process for a large friction force $F_f/a
\gamma_{12} = 0.07$.  The particles accumulate at the contact line and
introduce self-pinning of the contact line.  Finally, a ring-like deposit is
left.}
\label{fig:frb}
 \end{figure}

To understand the behavior of the contact line, we show the drying process of
the colloidal suspension droplet obtained in our simulations in~\figref{frb}.
The particles are initially randomly distributed in the droplet and some are
adsorbed at the droplet interface after equilibration.  During evaporation in
CR mode, the particles are transported to the contact line and accumulate there
(\figref{frb-1}).  At the end of the CR phase, the contact line intends to
recede and tries to pull the particles to move inward through a capillary
force. However, the particles are stuck at the substrate due to the large
friction and introduce a pinning of the contact line (\figref{frb-2}).  The
droplet further dries and more particles are transported to the contact line
forming a porous structure.  The interface recedes through the porous
structure, resulting in a slight decrease of the contact radius, as shown
in~\figref{bch-bigfri}.  Near the end of the lifetime of the droplet, the
contact angle decreases and the droplet flattens.  This thin flat film breaks
close to the ring (\figref{frb-3}), is dried rapidly, and finally a ring-like
deposit is left (\figref{frb-4}). 

\revisedtext{We note that the magnitude of the friction force $F_f/ a \gamma_{12}=0.07$ is chosen to be of the same order as the magnitude of the capillary force of the particles at the contact line, which has been calculated in the work of Sangani et al.~\cite{Sangani2009}. In reality, the friction force arises from the normal load on the particles and the surface roughness of both, particles and substrate. For a possible experimental validation of our results, we refer to recent atomic force microscopy measurements of the friction force between particles and a substrate~\cite{Guo2013}. }  

%% file: results-discussion.tex
\subsection{Theoretical model and discussion}
We propose a simple theoretical model to consider the effect of the friction force on
the effective radius of deposition.  In the case of a dilute colloidal
suspension droplet, we assume a uniform distribution of particles in the
droplet during evaporation.  Thus, the particle number density
$\rho_n=N_P/V_0$, where $N_P$ is the total number of particles, and $V_0$ is
the initial droplet volumer, keeps constant inside the evaporating droplet and
the remaining particles are transported to the contact line.  For simplicity,
we also assume that the particle deposition at the contact line is a monolayer
and the particles are uniformly distributed along the contact line.

\figref{fric-sketch} depicts the aggregation of particles at the contact line.
The outermost particle deforms the interface, thus a capillary force $F_c$
rises up and acts on the particle pointing inwards.  Meanwhile, the outermost
particle and its neighbouring particles along the radial direction experience a
friction force $F_f$, which resists the inward capillary force. We define the
outermost particle and its neighbouring particles along the radial direction as
a radial particle cluster.  The contact line depins when the capillary force is
larger than the total friction force $|F_c|\ge|F_{f}^{\mathrm{total}}|$, which
is given by $F_{f}^{\mathrm{total}} = N_{\mathrm{cluster}} F_{f}$.  Here,
$N_{\mathrm{cluster}}$ is the number of particles in the radial cluster.

\input{draw-md-fric.tex}
During the constant radius phase of evaporation, the contact angle of the
droplet decreases and the capillary force acting on the outermost particles
increases.  Meanwhile, more particles are transported to the contact line and
accumulate there resulting in an increase of the total friction force.
Therefore, the competition between capillary force and total friction force
determines the moment of depinning of the contact line and the deposition
pattern.

If the accumulated particles are not able to pin the contact line at the end of
the constant radius phase ($|F_c|\ge|F_{f}^{\mathrm{total}}|$), the contact
line depins and shrinks while the contact angle keeps constant. We assume the
capillary force acting on the outermost particles to be constant. However, the
total friction force increases due to the continuous accumulation of more
particles at the contact line. When the total friction force becomes larger
than the capillary force, the contact line is pinned at a position with a
certain distance $R_p$ to the center of the hydrophilic area.  The droplet
continues to dry in constant radius mode and leaves a deposition pattern with
an effective radius $R_p$ determined by 
\begin{equation}
 N_{\mathrm{cluster}}F_f=F_c
 \mbox{.}
 \label{eq:nfc}
\end{equation}

In order to derive an expression for $N_{\mathrm{cluster}}$, we assume that the
droplet keeps a spherical cap with volume $V_p$ when the contact line is pinned
at the position with distance $R_p$ to the center.  The total number of
particles that are transported to the contact line is given by
$(V_0-V_p)\rho_n$.
Based on a geometric analysis, we obtain
\begin{eqnarray}
 V_0=\frac{\pi h_0}{6}(3R_{0}^{2}+h_{0}^{2}) \mbox{,} \nonumber \\
 V_p=\frac{\pi h_p}{6}(3R_{p}^{2}+h_p^2) \mbox{,}
 \label{eq:v0vd}
\end{eqnarray}
where $h_p$ is the maximal droplet height when the self-pinning happens.  With
the assumption that the particles form a circular ring (outer radius $R_p$,
inner radius $r_p$) at the contact line, we obtain its area as
\begin{equation}
\pi (R_{p}^{2} - r_{p}^{2})=  (V_0-V_p)\rho_n\pi a^2 
\mbox{.}
\label{eq:ring}
\end{equation}
The average number of particles in the radial particle cluster is then
\begin{equation}
N_{cluster}= \frac{R_p-r_p}{2a} 
\mbox{.}
\label{eq:cluster}
\end{equation}
Using~\eqnref{ring} and~\eqnref{cluster}, we can write~\eqnref{nfc} as 
\begin{equation}
  \frac{R_p- \sqrt{R_{p}^{2}-(V_0-V_p)\rho_n a^2}}{2a}F_f = F_c
  \mbox{.}
  \label{eq:nfc2}
\end{equation}
\revisedtext{Based on~\eqnref{nfc2}, we note that the number of particles $(V_0-V_p)\rho_n$ at the contact line directly affects the 
pinning of the contact line, which is consistent with the conclusion drawn in the work of Weon et al.~\cite{Weon2013} that 
the particle packing fraction at the contact line plays an important role in self-pinning.}

Here, we determine a critical value of the friction force $F_{f}^{cr}$ below
which the contact line depins at the end of the constant radius phase.  At the
end of the constant radius phase, the number of particles that are transported
to the contact line is $(V_0-V_1)\rho_n$, where $V_1 = \frac{\pi
R_0(1-\cos \theta_d)}{6\sin \theta_d} \left[3R_{0}^{2}+\left(\frac{R_0(1-\cos
\theta_d)}{\sin \theta_d}\right)^2\right]$. We then obtain the critical value
of the friction force as
\begin{equation}
  F_{f}^{cr} = \frac{2aF_c}{R_0- \sqrt{R_{0}^{2}-(V_0-V_1)\rho_n a^2}}
  \mbox{.}
  \label{eq:nfcr}
\end{equation}
If the friction force $F_f<F_{f}^{cr}$, the contact line depins and follows a constant angle mode.
The contact angle of the droplet is $\theta_{d}$ and we can write 
$h_p=\frac{R_p}{\sin \theta_d}-\frac{R_p}{\tan \theta_d}= \frac{R_p(1-\cos \theta_d)}{\sin \theta_d}$.
Inserting~\eqnref{v0vd} into~\eqnref{nfc2} and after some manipulations we obtain 
\begin{widetext}
\begin{equation}
 R_p- \sqrt{R_{p}^{2}+\frac{\pi \rho_n a^2(1-\cos \theta_d)}{6\sin \theta_d}  
 \left( 3+\frac{(1-\cos \theta_d)^2}{(\sin \theta_d)^2}\right)
 R_{p}^{3}- N_p a^2} = 2 a F_c/F_f 
 \mbox{.}
\label{eq:nfcf}
 \end{equation}
 \end{widetext}

We note that there should be a minimal radius of deposition where all particles
are located around the center of the substrate.  This minimal radius can be
obtained as $R^{\mathrm{mi}}_p= \sqrt{N_p a^2} \approx 67$ and allows to
estimate the minimal friction force $F^{mi}_f$ below which particles will
always be located around the center as 
\begin{equation}
 F_c/F^{mi}_f = R^{\mathrm{mi}}_p/2a
 \mbox{.}
 \label{eq:fric_min}
\end{equation}

To validate our theoretical analysis, we carry out simulations 
with different friction forces in the range $0 \le F_f/a\gamma_{12} \le 0.07$. 
We define the effective radius $R_p$ of deposition in a way such that 
a circle of radius $R_p$ covers $95\%$ of all particles in the system.
\begin{figure}[h!]
\centering
\includegraphics[width= 0.4\textwidth]{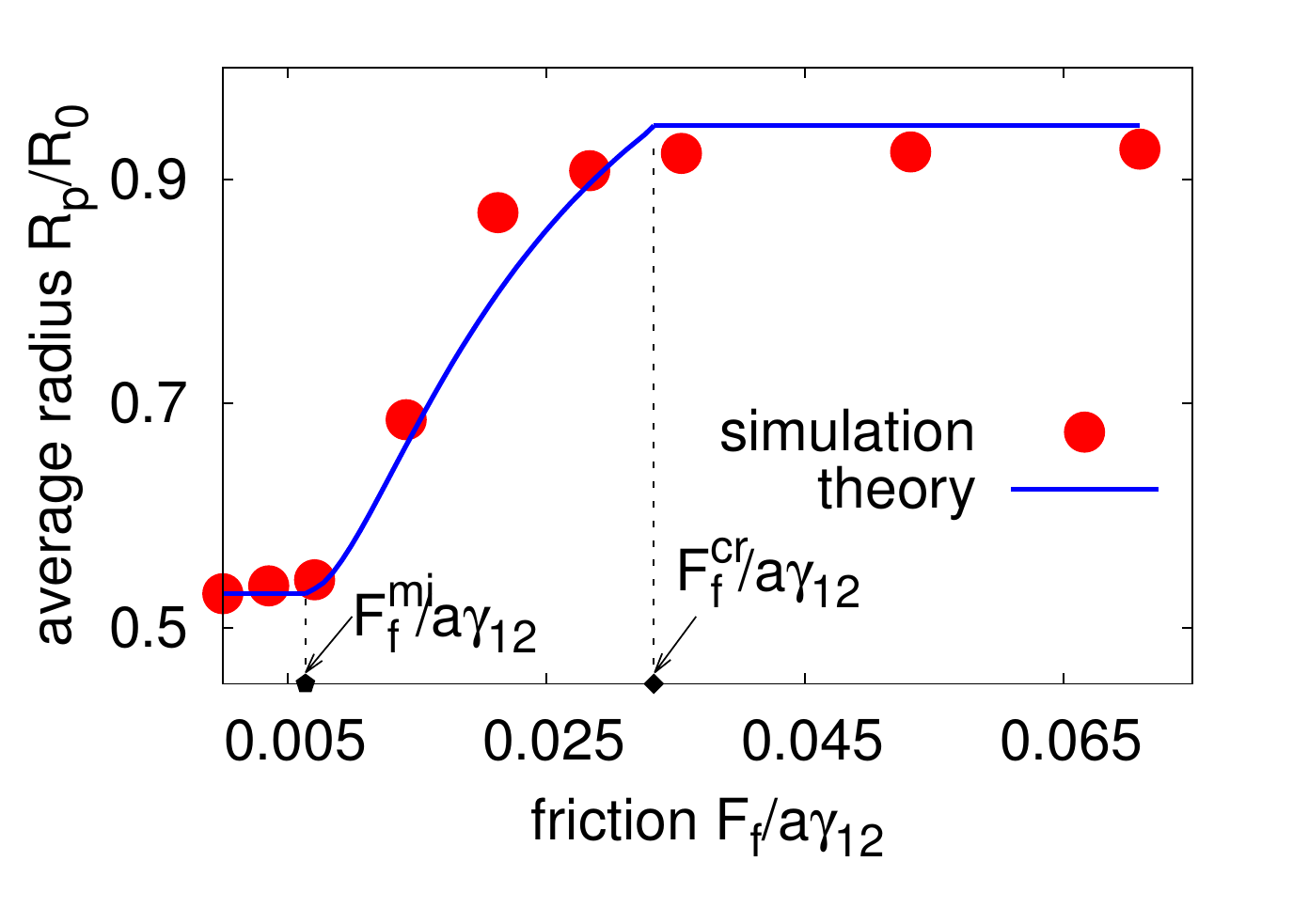}
\caption{The normalized effective radius $R_p/R_0$ of deposition as a function of the friction force $F_f/a\gamma_{12}$.
The theoretical analysis (Eq.($21-23$)) agrees well with our simulation data (symbols).} 
\label{fig:fric-radius}
\end{figure}
We solve~\eqnref{nfcr},~\eqnref{nfcf} and~\eqnref{fric_min} with the simulation
parameters $N_p=500$, $a=3$, $\theta_d=56^{\circ}$, $R_0=115$, $h_0=115$.  The
capillary force is assumed to be constant and we estimate it as $F_c/a
\gamma_{12} \approx 0.068$~\cite{Sangani2009}.  From~\eqnref{nfcr}
and~\eqnref{fric_min} we obtain the critical friction force $F_{f}^{cr}/a
\gamma_{12} \approx 0.034$ and the minimum firction force $F_{f}^{mi}/a
\gamma_{12} \approx 0.0064$, respectively.  In~\figref{fric-radius} we compare
the simulation results (symbols) with our theoretical analysis (solid lines).
The simulations show that the effective radius of deposition $R_p$ keeps
constant until the friction force reaches the minimal friction force
$F^{mi}_f$. Then, the effective radius increases for small friction forces, and
finally reaches a plateau for large friction forces which is in a good
agreement with our theoretical analysis. We note that the theoretical analysis
predicts a lower value of the effective radius than simulation results at
$F_{f}/a \gamma_{12} \approx 0.022$.  A possible interpretation is that near
the end of the lifetime of the droplet, while the particles are being pulled
inwards, the droplet flattens, breaks up and dries quickly. Therefore, the
capillary force vanishes and the particles stop moving, resulting a slightly
higher effective radius than expected from the theoretical analysis.

\revisedtext{We note that our theoretical analysis is only valid when the particles form a monolayer at the contact line, as observed in the experiments by Sangani at al.~\cite{Sangani2009}. However, the multilayer structures are also observed in evaporation-driven deposition~\cite{Abkarian2004,Marin2011}. Our theory can be extended to suit for multilayer cases with the structure of the multilayer 
being provided, such as the thickness of the multilayer along the radial direction.}

%% file: draw-md-fric.tex
\begin{figure}[!h]
\centering
 \begin{tikzpicture}
	
 \draw[ultra thick, draw=white!50!black] (-3.0,0.0) to (3.0,0.0);
\draw[red] (3.0,0.0) to (3.0,0.05);
	
\draw[ultra thick,red] (-2.75,0.0)  to  [out=30,in=-165](3,2.6);

       \filldraw[fill=green!80!black, draw=green!80!black]
      (-1.95,0.28) -- (-1.7,0.28) arc (0:360:0.25) -- (-1.70,0.28);
       \filldraw[fill=green!80!black, draw=green!80!black]
      (-1.45,0.28) -- (-1.20,0.28) arc (0:360:0.25) -- (-1.20,0.28);
        \filldraw[fill=green!80!black, draw=green!80!black]
      (-0.95,0.28) -- (-0.70,0.28) arc (0:360:0.25) -- (-0.70,0.28);

 \draw[thick, black,arrows=->] (-1.95,0.25) to (-1.30,0.25);
 \node at (-2.2,0.8) [fontscale=2] {$F_{c}$};
   \draw[thick, black,arrows=->] (-1.95,0.09) to (-2.2,0.09);
 \node at (-2.2,-0.28) [fontscale=2] {$F_{f_1}$};
  \draw[thick, black,arrows=->] (-1.45,0.09) to (-1.7,0.09);
 \node at (-1.5,-0.28) [fontscale=2] {$F_{f_2}$};
       \draw[thick, black,arrows=->] (-0.95,0.09) to (-1.2,0.09);
 \node at (-0.8,-0.28) [fontscale=2] {$F_{f_3}$};
 
  \filldraw[fill=green!80!black, draw=green!80!black]
      (1.50,1.0) -- (1.75,1.0) arc (0:360:0.25) -- (1.50,1.0);
  
     \filldraw[fill=green!80!black, draw=green!80!black]
      (2.0,1.9) -- (2.25,1.9) arc (0:360:0.25) -- (2.0,1.9); 
      
         \filldraw[fill=green!80!black, draw=green!80!black]
      (0.2,0.75) -- (0.2,0.75) arc (0:360:0.25) -- (0.2,0.75); 
 
\end{tikzpicture}
\caption{Sketch of the accumulation of colloidal particles at the contact line. The red line represents the droplet interface. 
The outermost particle experiences an inward capillary force $F_{c}$. 
The three particles in the radial particle cluster are subjected to the friction 
force $F_{f1}$, $F_{f2}$, and $F_{f3}$, respectively. 
When $F_{c}>F_{f1}+F_{f1}+F_{f3}$, the contact line will depin, 
and when $F_c<F_{f1}+F_{f2}+F_{f3}$, the contact line will be pinned by the particles.}
\label{fig:fric-sketch}
\end{figure}

%% file: conclusion.tex
\section{Conclusion}
\label{sec:final}
We investigated the deposition of a drying droplet on a chemically patterned
substrate and demonstrate that we are able to control the wettability of the
substrate accurately to reproduce the constant radius mode and the stick-slip
mode. When the droplet evaporates with a pinned contact angle, the radial
velocity inside the droplet diverges near the end of the droplet lifetime,
which is in a good agreement with theory and experiment.

We then studied the effect of friction between the particles and the substrate
on the contact line dynamics and the final deposition pattern in drying
colloidal suspension droplets.  Without friction, the contact line shows a
stick-slip drying mode similar to a drying pure droplet.  Interestingly, due to
the pinning of the contact line on the particles, the droplet in the slip stage
has a larger contact angle than that of a pure droplet.  Moreover, the
particles follow the receding contact line and finally form a dot-like
deposit.

With increasing the friction force, the drying suspension droplet shows a
transition from stick-slip mode to constant radius mode. With a large friction
force, the particles accumulate at the contact line, introducing self-pinning.
Thus, more particles are transported to the contact line and form porous
structures.  Surprisingly, we find that the contact line recedes through the
pores, resulting in a slight decrease of the contact radius. 

\revisedtext{Moreover}, we propose a theoretical model for the effective radius of deposition
as a function of the friction force. Our model predicts a critical friction
force at which the self-pinning happens and a decreasing effective radius of
the deposit with decreasing friction. Additionally, the effective radius of the
particle deposit keeps constant for all friction forces smaller than a minimal
friction. We carried out simulations and find good agreement between the
simulation results and our theoretical predictions.  Our results contribute to
the fundamental understanding of self-pinning phenomena and can find
implications for developing active control strategies for the deposition of
drying droplets.

\revisedtext{Finally, we note that after the pioneering work of Deegan et al. 
on the coffee-ring effect~\cite{Deegan1997}, 
the understanding of the macroscopic mass and momentum transport phenomena in 
drying droplets is relatively well developed~\cite{Deegan1997, Hu2005b, Popov2005, Detlef2015}.  
However, the interfacial phenomena, contact-line and deposition processes in drying colloidal 
suspension droplets are not nearly as well  understood~\cite{Larson2014}.  
The available theoretical models usually apply a convection-diffusion equation 
governing the transport of colloidal particles~\cite{Fleck2015,Pham2017}, which cannot 
account for phenomena at the scale of individual particles 
(e.g., contact-line pinning on particles, particle-particle interactions, 
particle-substrate interactions). We perform lattice Boltzmann simulations with fully resolved colloidal particles. Thus, 
we are able to fully resolve the information at the scale of individual particles and 
can gain more insight in the deposition process.}

%% file: ring-main-supple.tex
\section{Derivation of the radial fluid velocity in a drying droplet with pinned contact line}
The average radial velocity can be derived assuming mass conservation~\cite{Marin2011}.
We consider a small volume $\Delta v$. The mass conservation law of this controlled volume $\Delta v$ can be written as
\begin{equation}
 \frac{\partial h(r,t)}{\partial t} = -\frac{1}{r}\frac{\partial}{\partial r} Q(r,t) - \frac{1}{\rho} J(r,t)
 \mbox{,}
 \label{eq:masscona}
\end{equation}
where $h(r,t)$ is the local height of the droplet surface, $t$ is the time, $r$
is the radial coordinate, $Q(r, t)$ is the volume flow, $\rho$ is the liquid
density, and $J(r, t)$ is the evaporative flux.

In the limit of small contact angles, the evaporative flux can be written
as~\cite{Deegan1997, Popov2005, Marin2011}
\begin{equation}
J(r)= \frac{2}{\pi} \frac{D\Delta \rho}{\sqrt{R_{d}^{2}-r^2}},
\label{eq:fluxa}
\end{equation}
with $D$ the diffusion constant for the droplet's liquid in another fluid, and
$\Delta \rho$ the difference between the liquid density just above the drop
surface and the liquid density located far away from its surface.
\revisedtext{Here, we assume that the droplet has spherical cap shape and we obtain $h(r,t) = \frac{R_{d}^{2}-r^2}{R_{d}^{2}}h(0,t)$. 
We note that LBM can recovers Navier-stokes equation~\cite{Succi2001}, 
thus we are able to simulate a droplet with larger contact angle and a 
spherical-cap shape, which goes beyond the limitation of lubrication theory 
based models of drying droplets~\cite{DIDDENS2017}, 
where an assumption of a flat droplet is required.}

Following global mass conservation, the change in droplet volume 
\begin{equation}
 \frac{dV}{dt} = \frac{d}{dt} \int_{o}^{R_d} h(r,t)2\pi r dr = \frac{\pi R_{d}^{2}}{2} \frac{dh(0,t)}{dt}
\label{eq:vmassa}
 \end{equation}
should be equal to the total amount of evaporated liquid
\begin{equation}
 \frac{dV}{dt} = \frac{-1}{\rho} \int_{0}^{R_d} J(r)2\pi r dr = \frac{-4 R_d D \Delta \rho}{\rho}
 \mbox{.}
\label{eq:vfluxa}
 \end{equation}
Comparing~\eqnref{vmassa}, and~\eqnref{vfluxa}, we obtain 
\begin{equation}
 \frac{dh(0,t)}{dt} = \frac{-8D \Delta \rho}{\pi \rho R_d}
 \mbox{.}
\end{equation}
The evolution of the droplet height is
\begin{equation}
 h(0,t) =  \frac{8D \Delta \rho}{\pi \rho R_d} (t_e-t)
 \mbox{,}
 \label{eq:htimea}
\end{equation}
with $t_e$ being the total life time of the droplet. We note that~\eqnref{htimea} predicts 
that the droplet height decreases linearly with time.
 
Inserting~\eqnref{htimea} and~\eqnref{fluxa} into~\eqnref{masscona} and 
after some manipulations we obtain
\begin{equation}
 Q(r) = \frac{2D \Delta \rho}{\pi \rho} \left[ \sqrt{R_{d}^{2}-r^2} - \frac{1}{R_{d}^{3}}(R_{d}^{2}-r^2)^2 \right]
 \mbox{.}
 \label{eq:volumeflowa}
\end{equation}
The volume flow $Q(r)= rh \bar{u}$, where $\bar{u}$ is the height averaged radial velocity.
Using~\eqnref{volumeflowa} and~\eqnref{htimea}, it can be derived as
\begin{eqnarray}
\!\!\!\!\!\!\!\!\!\!\!\!\!\! \bar{u} &=& \frac{Q}{rh(r,t)} \nonumber \\
 &=& \frac{2R_{d}^{2}D\Delta \rho}{\pi \rho r} \frac{1}{h(0,t)} \left[ \frac{1}{\sqrt{R_{d}^{2}-r^2}}  - \frac{1}{R_{d}^{3}}(R_{d}^{2}-r^2) \right]\nonumber
 \mbox{.}
 \\
 \label{eqn:velavea}
\end{eqnarray}
Following an approach proposed by Marin et al.~\cite{Marin2011} based 
on the lubrication approximation, we can then estimate the radial velocity as 
\begin{equation}
 u(r,z,t) = \frac{3}{h^2(r,t)}  \bar{u} \left(h(r,t)z-\frac{1}{2}z^2 \right)
 \mbox{.}
 \label{eq:radiava}
\end{equation}